\documentclass[12pt,preprint]{aastex}

\newcommand{\etal}{et al.}

\newcommand{\msun}{M$_{\sun}$\,}

\newcommand{\ha}{H$\alpha$}

\shorttitle{IFS LGS-AO of a Candidate Disk Galaxy at z$\sim$1.5}

\shortauthors{Wright, S.~A. \etal}

\begin{document}

\title{Integral Field Spectroscopy of a Candidate Disk Galaxy at z$\sim$1.5 using Laser Guide Star Adaptive Optics}

\author{S.~A. Wright\altaffilmark{1},  J.~E. Larkin\altaffilmark{1},
  M.~Barczys\altaffilmark{1}, D.~K. Erb\altaffilmark{2}, C.~Iserlohe\altaffilmark{3},
  A.~Krabbe\altaffilmark{3},  D.~R. Law\altaffilmark{4},
  M.~W. McElwain\altaffilmark{1}, A.~Quirrenbach\altaffilmark{5},
  C.~C. Steidel\altaffilmark{4}, J.~Weiss\altaffilmark{1}}   

\begin{abstract}

We present 0.$\arcsec$1 resolution near-infrared integral field spectroscopy of \ha~in a z=1.4781 star forming galaxy, Q2343-BM133. These observations were obtained with OSIRIS (OH Suppressing Infra-Red Imaging Spectrograph) using the W.M. Keck Observatory Laser Guide Star Adaptive Optics system. \ha~emission is resolved over a 0\farcs8 (6.8 kpc) x 0\farcs5 (4.3 kpc) region with a 0\farcs1 spatial resolution. We find a global flux of 4.2$\pm$0.6$\times$10$^{-16}$ ergs s$^{-1}$ cm$^{-2}$, and detect a spatially resolved velocity gradient of $\sim$134 km s$^{-1}$ across the galaxy and a global velocity dispersion of 73$\pm$9 km s$^{-1}$. An upper limit of [N{\sc ii}]/\ha $\lesssim$ 0.12 is inferred, which implies that this galaxy is not dominated by an active galactic nucleus and has a metallicity at or below 1/2 solar metallicity. We derive a star formation rate (SFR) of 47$\pm$6 \msun yr$^{-1}$, and a dereddened SFR of 66$\pm$9 \msun yr$^{-1}$. Two-dimensional kinematics for Q2343-BM133 fit well with an inclined-disk model, with which we estimate an enclosed mass of 4.3$\times$10$^{9}$ \msun within 5.5 kpc. A possible merger scenario is also presented, and can not be fully ruled out. We derive a virial mass of 1.1$\times$10$^{10}$ \msun for a disk geometry, using the observed velocity dispersion. We propose that Q2343-BM133 is currently at an early stage of disk formation at a look-back time of 9.3 Gyr.

\end{abstract}


\altaffiltext{1}{Department of Physics and Astronomy,
University of California, Los Angeles, CA 90095;
saw, larkin, barczysm, mcelwain, weiss@astro.ucla.edu}
\altaffiltext{2}{Harvard-Smithsonian Center 
of Astrophysics, MS20, 60 Garden St, Cambridge, MA 02138;
derb@cfa.harvard.edu}
\altaffiltext{3}{Institute of Physics, 
University of Cologne, 50937, Germany; 
iserlohe, krabbe@ph1.uni-koeln.de}
\altaffiltext{4}{California Institute of Technology, 
MS 105-24, Pasadena, CA 91125; 
drlaw, ccs@astro.caltech.edu}
\altaffiltext{5}{ZAH Landessternwarte, Koenigstuhl,
D-69117 Heidelberg, Germany; 
A.Quirrenbach@lsw.uniheidelberg.de} 

\section{Introduction}\label{intro}

Formation and evolution of galactic disks remains one of the most important unsolved aspects of extragalactic study. At redshifts up to z $\sim$ 1, optical spectroscopy and Hubble Space Telescope (HST) imaging have produced a wealth of data on disk galaxies (eg, \citealt{vogt96}, \citealt{vogt97}, \citealt{lilly98}).  However, at higher redshifts galaxy morphologies become difficult to study since their bolometric surface brightnesses fade as (1+z)$^{4}$ and rapidly shrink to angular scales under one arcsecond. Beyond a redshift of one, optical observations are only able to detect rest frame ultraviolet (UV) light and older stellar populations can be overlooked. Because of these observational limitations, spatial densities and properties of disk galaxies at higher redshift (z $\ga$ 1) are still unknown, and we have little knowledge about the epoch of disk formation.  

With recent photometric and spectroscopic deep surveys our understanding of high redshift (1.4 $\le$ z $\le$ 3.5) star-forming galaxies has greatly increased, yielding properties of mass, stellar populations, gas and dust content, and star formation rates. Most successful among these, are color selection techniques including the search for galaxies with a strong Lyman Break between optical imaging filters. These Lyman Break Galaxies (LBGs) are rapidly star forming systems that are precursors to some of the largest modern galaxies \citep{steidel96, madau96}. They provide an excellent sample for observing the early development of structures of star forming galaxies. To date, LBGs have primarily been studied in rest-frame UV light to investigate large-scale clustering, global star formation rates (SFR), and production of metals and outflows. HST observations of LBGs have shown structure down to 8 kpc in size, which is usually clumpy and irregular and is indicative of regions of active star formation \citep{consel03,lotz06,law06}. 

There has been growing evidence which suggests that at least a subset of these LBG systems have disk-like morphology at rest-frame optical wavelengths (\citealt{lab03}) and show large organized rotation which may indicate formation of an early galactic disk (\citealt{van01}, \citealt{erb03}). These studies have used NIR slit-spectroscopy of \ha~emission under seeing-limited conditions. At these redshifts, the best seeing-limited observations probe roughly 5 kpc and yield only 2 or 3 resolution elements across the typical galaxy. Many have argued that these kinematic measurements are insufficient to claim rotation since there are only three independent samples across the galaxies, and their curves could easily be duplicated with a variety of clumpy distributions. 

With the advent of integral field spectrographs (IFS), observations are now able to extend this study to two-dimensional (2D) mapping of galaxy's kinematics and morphology. Recently, the IFS SINFONI (VLT) was used to conduct seeing-limited observations of 14 galaxies at z$\sim$2 \citep{schreib06}, which showed four galaxies that exhibit rotating disks with clumpy morphologies. However, the \citet{schreib06} study also observed very high internal velocity dispersions (often $\gtrsim$ 100 km/sec) and concluded that their disks may be short lived and not precursors of modern disk galaxies. Again, with a seeing limited spatial resolution it is difficult to reliably interpret kinematics in objects that are roughly an arcsecond in diameter. A star forming galaxy has now also been imaged with SINFONI using a natural guide star adaptive optics (AO) system. This observation \citep{genzel06} suggest that the z=2.38 galaxy has a massive rotating progenitor disk. However, their best fit model has residual velocities comparable to the original measurements, indicating that significant deviations from simple rotation are present. These IFS studies have targeted z $\gtrsim$ 2 galaxies, with a look-back time of more than 10 Gyr, which is greater than the expected formation age of the Milky Way disk. 

Using Keck Observatory's Laser Guide Star Adaptive Optics (LGS-AO) and the IFS of OSIRIS (OH Suppressing Infra-Red Imaging Spectrograph) we have obtained a spatially resolved ($\lesssim$ 0\farcs1) spectrum of \ha~emission from the z=1.4781 star forming galaxy Q2343-BM133. OSIRIS coupled with LGS-AO allows full two-dimensional mapping of \ha~morphology and dynamics of the galaxy. We selected Q2343-BM133 as one of our targets from the rest-UV color selected catalog of \citet{steidel03,steidel04,adel04}. Q2343-BM133 was chosen for its compactness, and because it is one of the bluest objects in the Steidel catalog with an R-K color of 2.1. In \S\ref{observ} we describe the observations and reductions. In \S\ref{results} we present general Q2343-BM133 morphology and kinematics, and in \S\ref{disc} we discuss \ha~kinematics, disk modeling, mass distribution, metallicity, and a possible merger scenario. In this paper we assume $\Lambda$-dominated cosmology with $\Omega_{M}$=0.27, $\Omega_{V}$=0.73, and H$_{o}$=71 km sec$^{-1}$ Mpc$^{-1}$ \citep{benn03}. For our redshift of z=1.4781 this corresponds to $\simeq$ 8.53 kpc per arcsecond and a luminosity distance of 10.8 Gpc.

\section{Observations \& Data Reduction}\label{observ}

Q2343-BM133 (hereafter: BM133) was the first successful detection of a high-redshift star forming galaxy using the new integral field spectrograph OSIRIS \citep{larkin06}. We obtained data on BM133 with OSIRIS in conjunction with the Keck LGS-AO system \citep{wiz06,mvdam06}, as part of commissioning time on November 23, 2005. The OSIRIS spectrograph utilizes a lenslet array, grating, and 2048x2048 Rockwell Hawaii II HgCdTe infrared detector to obtain simultaneous spectra over a rectangular field of view at a spectral resolution of R$\sim$3000. Reimaging optics in OSIRIS allow the user to change the spatial resolution on the lenslet array. We observed BM133 in the Hn3 ($\lambda_{cen}$ = 1.6348$\micron$, $\Delta\lambda$ = 0.0814 $\micron$) filter with a spatial sampling of 0\farcs1 in both lenslet axes, which provides a 4\farcs8 x 6\farcs4 field of view. 

The LGS-AO system produced an artifical guide star directly on the OSIRIS optical axis, with the tip-tilt sensor closed on an R = 16.2 mag star located 55\farcs2 away from BM133. We chose an instrument position angle of 90$^\circ$ to maintain the science target and tip-tilt star within the AO and OSIRIS fields of view, respectively. To ensure acquisition and assess the quality of the night we observed the tip-tilt star with OSIRIS LGS-AO before moving to the galaxy, and measured the point spread function full-width at half-maximum (FWHM) of $\lesssim$ 0\farcs1. This is an upper limit to the optical system performance since we have undersampled the PSF of the AO correction at this lenslet resolution. For the given magnitude and separation of the tip-tilt star, the Keck LGS-AO system predicts a Strehl ratio of 24\%. For each dataset we took a 900 second frame on source with BM133 shifted 1\arcsec~north of the field center, followed by a 5\arcsec~eastern dither for a 900 second sky exposure, and then another on source 900 second exposure with BM133 shifted 1\arcsec~south of the field center. We repeated this data set three times, yielding a total exposure of 90 minutes on BM133 with three 900 second sky exposures to ensure good quality sky subtraction.  

Reductions were performed using the OSIRIS final data reduction pipeline (DRP) in combination with a custom IDL emission-line extraction routine.  The DRP reduction steps used are similar to those performed for other NIR spectrograph instruments: correction of detector non-linearity, removal of bad pixels, removal of detector cross talk, sky subtraction, wavelength calibration, and mosaicing of images. Wavelength calibration is determined uniquely for each of the resulting 3000 spectra using arc line lamps. The critical and unique reduction steps for this data set are extraction of emission line spectra and assembly of the 3D cube (x and y spatial locations of lenslets and $\lambda$ wavelength), in which the flux for each lenslet is the integral of the peak pixel values for each wavelength channel. 

\section{Results}\label{results}

\begin{figure*}[t]
\epsscale{1.0}
\begin{center}
\plotone{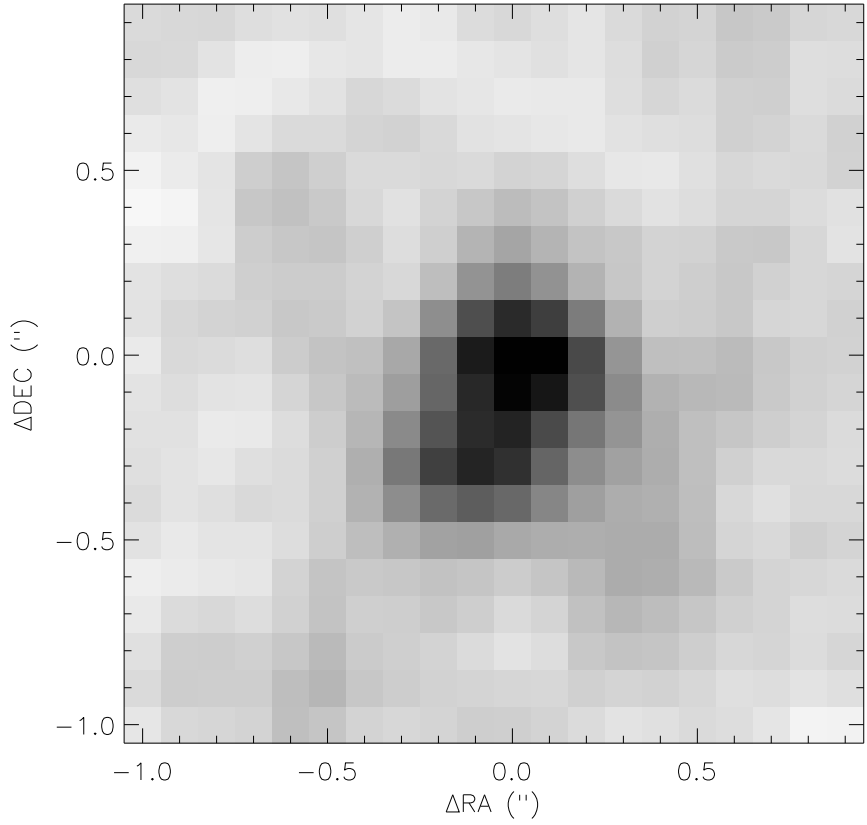}
\caption{A Gaussian smoothed (FWHM=0\farcs2) mosaiced image of Q2343-BM133 with a total exposure of 90 minutes collapsed around H$\alpha$ ($\Delta$$\lambda$ = 1.4 nm) with a spatial size of 2\farcs0$\times$2$\farcs$0 and position angle of 0$^{\circ}$.\label{bm133}}
\end{center}
\end{figure*}

The integrated \ha~flux was collapsed from the final 90 minute mosaic frame using an average over a bandpass of $\Delta\lambda$ = 0.014 $\micron$, which is presented in Figure \ref{bm133}. The final mosaiced image has been smoothed spatially with a Gaussian kernel (FWHM 0\farcs2) to achieve higher signal-to-noise for our kinematic analysis. The \ha~flux emission of BM133 is resolved spatially and extends over a 0\farcs8 (6.8 kpc) $\times$ 0.\arcsec5 (4.3 kpc) region oriented from the northwest (NW) to the southeast (SE). We define the center of BM133 as the dynamical center as determined in \S\ref{disc}, and use this position for all figures. The morphology of BM133 is suggestive of a disk-like galaxy with \ha~brightest in the NW region, and the \ha~flux tailing off towards the SE region of the galaxy. This may indicate that the presence of dust extinction and/or the distribution of \ha~star-forming (SF) regions are not uniform throughout the galaxy, but may have a more clumpy distribution as seen in other high-z star forming galaxies \citep{papo05,schreib06,genzel06}. 
 
\begin{figure*}[t]
\epsscale{1.0}
\begin{center}
\plotone{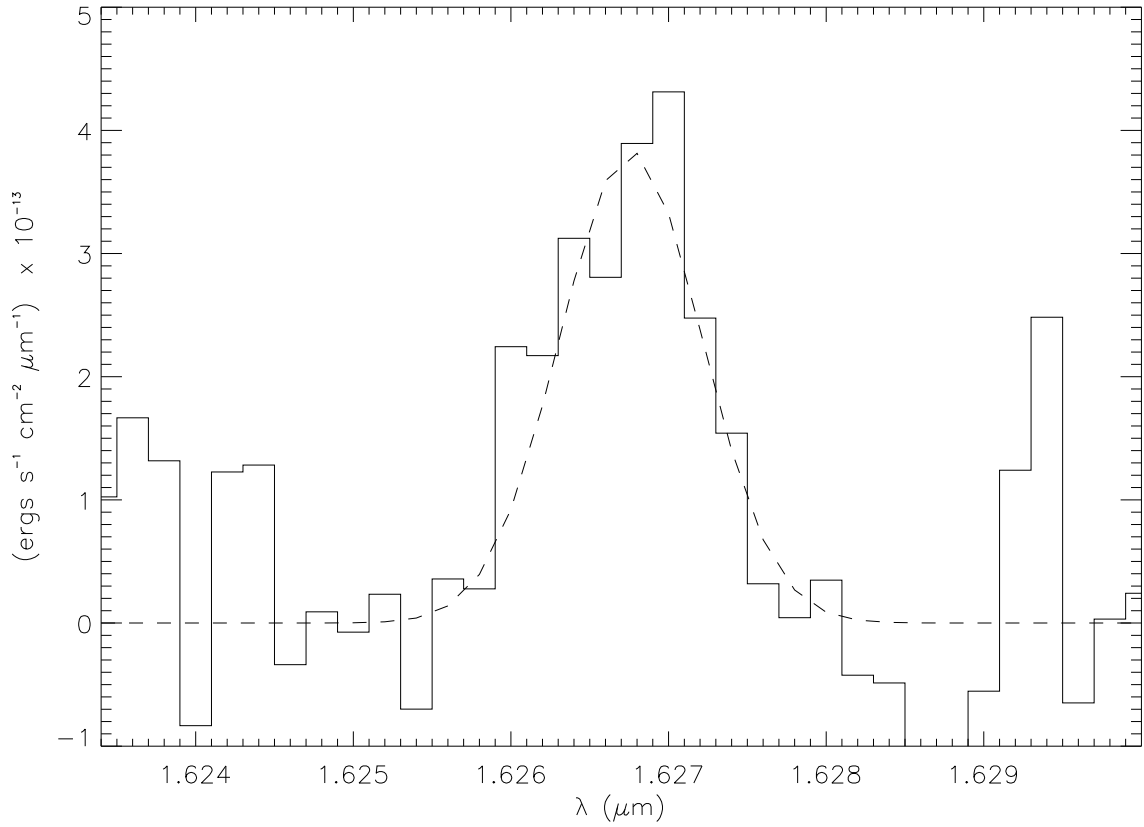}
\caption{Total \ha~emission spectrum of Q2343-BM133, collapsed over a spatial size of 1\farcs3$\times$1\farcs3 encompassing the entire extent of the galaxy. A Gaussian is fitted over the \ha~emission to yield our global velocity dispersion ($\sigma_{v}$) with intrumental correction performed.\label{spec}}
\end{center}
\end{figure*}

\ha~emission from BM133 has previously been detected with NIRSPEC with a derived H$\alpha$ flux of 3$\times$10$^{-16}$ ergs s$^{-1}$ cm$^{-2}$ \citep{erb06a,erb06b}. This was a seeing-limited slit spectroscopy observation, which yielded a flux measurement with no 2D spatial information. Figure \ref{spec} is the spectrum of the summation of the entire spatial extent (1\farcs3 $\times$ 1\farcs3) of the unsmoothed final mosaic frame of BM133, where we find a total integrated H$\alpha$ flux of 4.2$\pm$0.6$\times$10$^{-16}$ ergs s$^{-1}$ cm$^{-2}$. This is 40\% more flux than found by \citet{erb06a, erb06b}, but we were not biased by slit orientation and are able to integrate the flux across the entire extent of the galaxy. \ha~emission provides an instantaneous measurement of SF regardless of the global stellar population and history. We find an \ha~luminosity of 5.9x10$^{42}$ ergs s$^{-1}$, and using the Global Schmidt law (SFR (\msun yr$^{-1}$) = L$_{H_{\alpha}}$ / 1.26$\times$10$^{41}$ ergs s$^{-1}$: \citealt{kenn98}) we find a global SFR of 47$\pm$6 \msun yr$^{-1}$. \citealt{erb03} used optical colors and SED fitting to find an E(B-V)=0.115 for BM133. Applying this correction to the \ha~luminosity results in a dereddened SFR$_{dered}$= 66$\pm$9 \msun yr$^{-1}$. We did not detect [N{\sc ii}] ($\lambda_{rest}$ = 0.6548 and 0.6583 $\micron$) emission. However, using a square aperture of size 0\farcs5 centered on the \ha~flux location and the expected wavelengths of [N{\sc ii}] emission, we were able to set an upper limit of [N{\sc ii}]/\ha~$\lesssim$ 0.12.  

Fitting a Gaussian to the spatially integrated global spectrum of BM133 (Figure \ref{spec}) we find a systemic redshift of z=1.4781. The H$\alpha$ velocity width determined from the Gaussian width were corrected for the instrumental root-mean-square line width (43 km s$^{-1}$) using atmospheric OH lines at nearby wavelengths, to yield a total velocity dispersion ($\sigma_{v}$) of 73$\pm$9 km s$^{-1}$. This is larger by 25 km s$^{-1}$ compared to $\sigma_{v}$ previously observed by \citet{erb06a,erb06b} with slit spectroscopy, and again our difference may be explained since OSIRIS samples the full \ha~flux emission across the galaxy. This discrepancy might also be due to the difficulty to measure $\sigma_{v}$ since NIRSPEC has a factor of 2 lower spectral resolution. Table 1 summarizes these observational properties of BM133.

\begin{figure*}[t]
\epsscale{1.0}
\begin{center}
\plotone{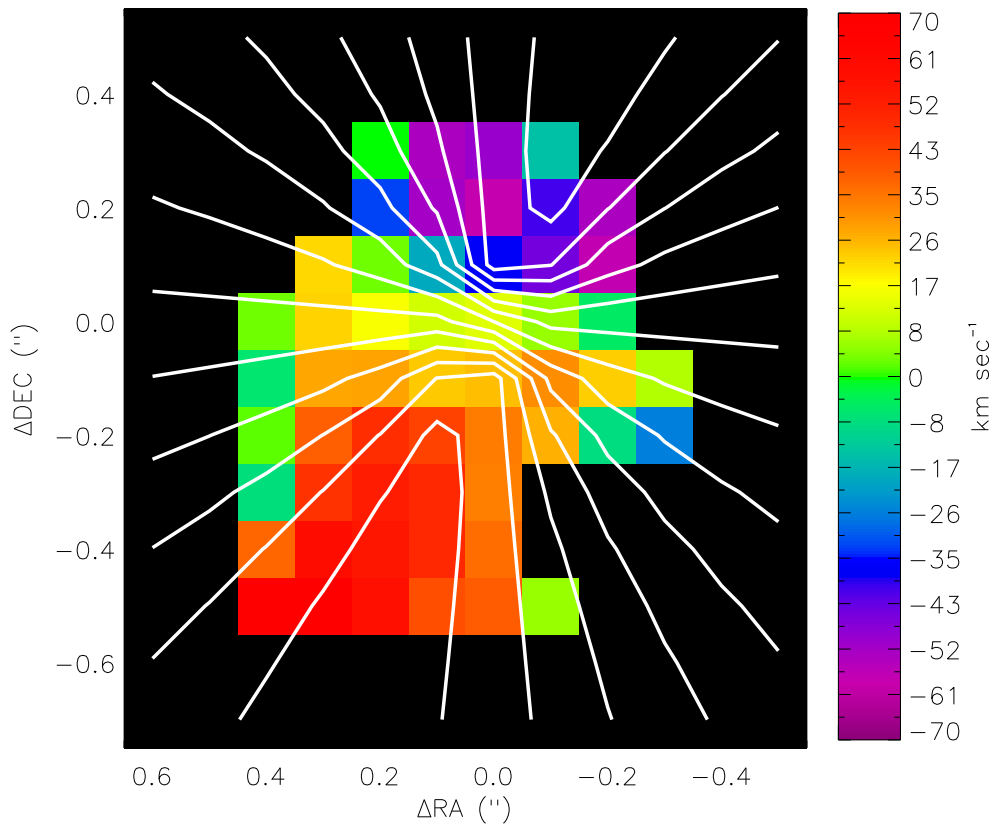}
\caption{Radial Velocity: Two dimensional \ha ~kinematics of Q2343-BM133 showing the spatial distribution of velocity (km s$^{-1}$) relative to the measured systemic velocity. This was obtained from Gaussian fits to each spatial element (lenslet) after smoothing spatially with a 2D Gaussian of FWHM=2 (or 0\farcs2). The velocities are color-coded matching the labeled color-bar. Overlaid is the spider diagram modeled in \S\ref{disc} with each contour representing 10 km s$^{-1}$.\label{velmap}}
\end{center}
\end{figure*}

\begin{figure*}[t]
\epsscale{1.0}
\begin{center}
\plotone{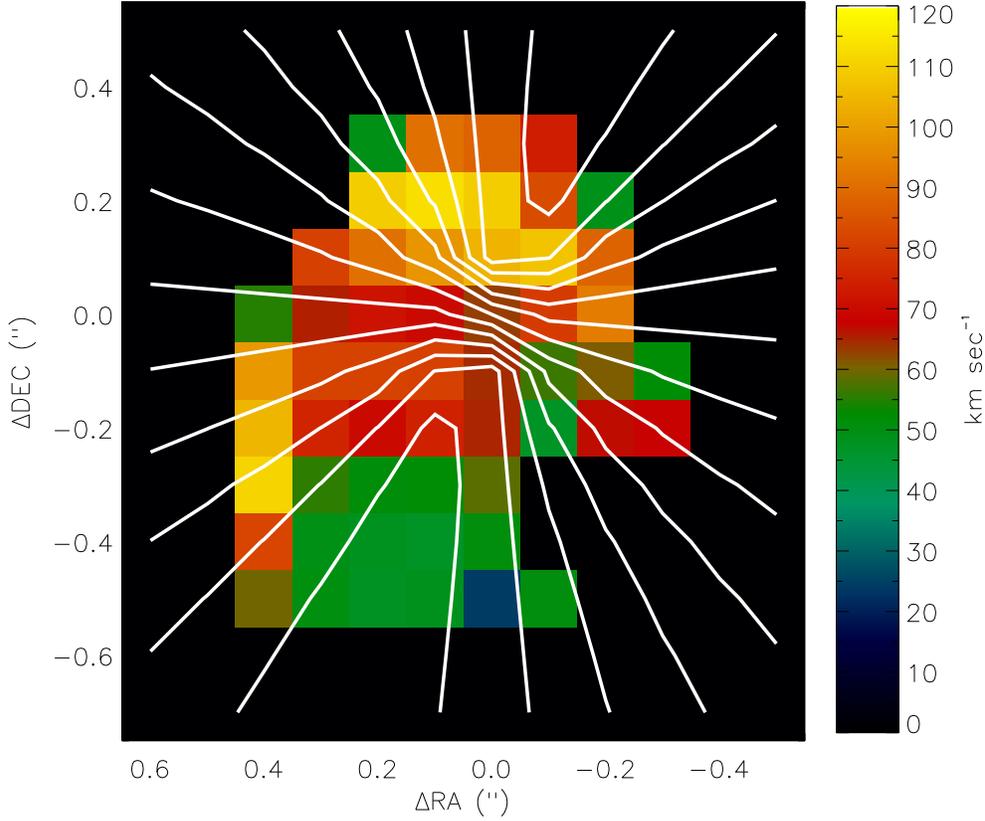}
\caption{Velocity Dispersion: Two dimensional \ha ~kinematics of Q2343-BM133 showing spatial distribution of velocity dispersion(km s$^{-1}$) after removal of instrumental widths. This was obtained from Gaussian fits to the data cubes after smoothing spatially with a two-dimensional Gaussian of FWHM=2 (or 0\farcs2). Velocities are color-coded according to the labeled color bar. Overlaid is the spider diagram modeled in \S\ref{disc} using the radial velocity offsets (see Figure \ref{velmap}). Note we see higher velocity dispersions in the NW region of the galaxy, nearest the dynamical center, and lower velocity dispersions nearest the plateau velocities, indicated by the overlaid spider diagram with each contour representing 10 km s$^{-1}$.\label{vdispmap}}
\end{center}
\end{figure*}

To measure the kinematic structure of BM133 we fit a Gaussian for each lenslet with an integrated signal-to-noise (S/N) $\gtrsim$ 20, yielding a velocity offset relative to the systemic redshift, and a velocity dispersion per lenslet. To account for nearby atmospheric OH sky line residuals, we determined weights for our Gaussian fits from the standard deviation per wavelength channel on a sky region within the 3D cube. Figure \ref{velmap} is a map of the velocity offsets at each lenslet, overlaid with a model spider diagram described in \S\ref{spider}. We find a velocity range of $\sim$134 km s$^{-1}$ across the galaxy, with negative velocity offsets concentrated in the NW region and positive velocities within the SE region. Note that the level of shear we see within BM133 would not have been detected in seeing-limited observations such as in the SINFONI dataset \citep{schreib06}. The H$\alpha$ velocity widths were also determined from the Gaussian fits at each lenslet location. The \ha~widths were then corrected for the instrumental line widths. These final line widths include the effects of beam smearing from the wings of the PSF and from the range of velocities that are spanned by a lenslet in the smoothed data cube (approximately 0\farcs2 lenslet$^{-1}$). Therefore, we consider these widths as upper limits to the velocity dispersions present at each spatial location. The resulting map of intrinsic velocity dispersions ($\sigma_v$) at each spatial location is presented in Figure \ref{vdispmap}, and are further discussed in \S\ref{spider}.

\section{Analysis \& Discussion}\label{disc}

\subsection{\ha~Kinematics}\label{spider}

The 2D velocity map for BM133 is indicative of a galaxy with a symmetrically rotating disk. This is seen in our velocity offsets with a velocity gradient across the major axis of BM133 and a steep, but continuous, velocity transition from the approaching and receding portions of the galaxy. We are able to map the line-of-sight radial velocities of BM133, but in order to know the true circular velocities of the galaxy we need to account for the orientation and projection effects. We have generated inclined-disk models \citep{bege87} and performed least squares fits to our observed velocity fields in order to find the appropriate disk model for BM133. The parameters in our model are rotation center of the galaxy (x$_{o}$,y$_{o}$), systemic velocity (V$_{o}$), position angle of the major axis ($\phi$), inclination angle (\textit{i}), and circular velocity (V$_{c}$(R)) as a function of radius from the center. We adopt a rotation curve profile that is commonly seen in local spiral galaxies, with a linearly rising circular velocity at lower radii and a plateau at larger radii. For V$_{c}$(R) we use two parameters for the fit using a slope (\textit{m}) and plateau radius (r$_{p}$) with the following equations,

\begin{equation}
V_{c} = mR, ~~~~~ \mbox{for R \textless~r$_{p}$} \\ 
\end{equation}
 
\begin{equation}
V_{c} = mr_{p}, ~~~~~ \mbox{for R $\gtrsim$~r$_{p}$} \\
\end{equation}

We modeled spider diagrams based on our seven parameters (x$_{o}$, y$_{o}$, V$_{o}$, \textit{i}, $\phi$, \textit{m}, r$_{p}$) and convolved our models with the same Gaussian kernel (FWHM = 0\farcs2) as our observed data. We performed a grid search over all parameters to find the global minimum reduced chi-squared ($\tilde{\chi}^{2}$). This minimum occurs at $\tilde{\chi}^{2}$ = 0.72 with the parameters: \textit{i}=54$^{\circ}$, $\phi$ = 160$^{\circ}$, r$_{p}$ = 0\farcs11 (0.9 kpc), and \textit{m} = 65 km s$^{-1}$ 0\farcs$1^{-1}$. The corresponding model spider diagram is overlaid on our velocity field in Figure \ref{velmap} with respect to the dynamical center of the galaxy. The center position (x$_{o}$,y$_{o}$), systemic velocity (V$_{o}$), and position angle ($\phi$) were tightly constrained on the observed fields, with a significant degeneracy of the model found with the inferred circular velocity and inclination. This is expected since V$_{c}$(R) and \textit{i} are correlated by V$_{c}$(R)sin(\textit{i}), and there is a known degeneracy between the two parameters that increases at larger inclinations \citep{bege87}. A second local $\tilde{\chi}^{2}$ minimum occurs at \textit{i} = 86$^{\circ}$. However, based on the ellipticity of the major and minor axes we rule out the second inclination (\textit{i} = 86$^{\circ}$) since the morphology is inconsistent with an edge-on galaxy.
 
\begin{figure*}[t]
\epsscale{1.0}
\begin{center}
\plotone{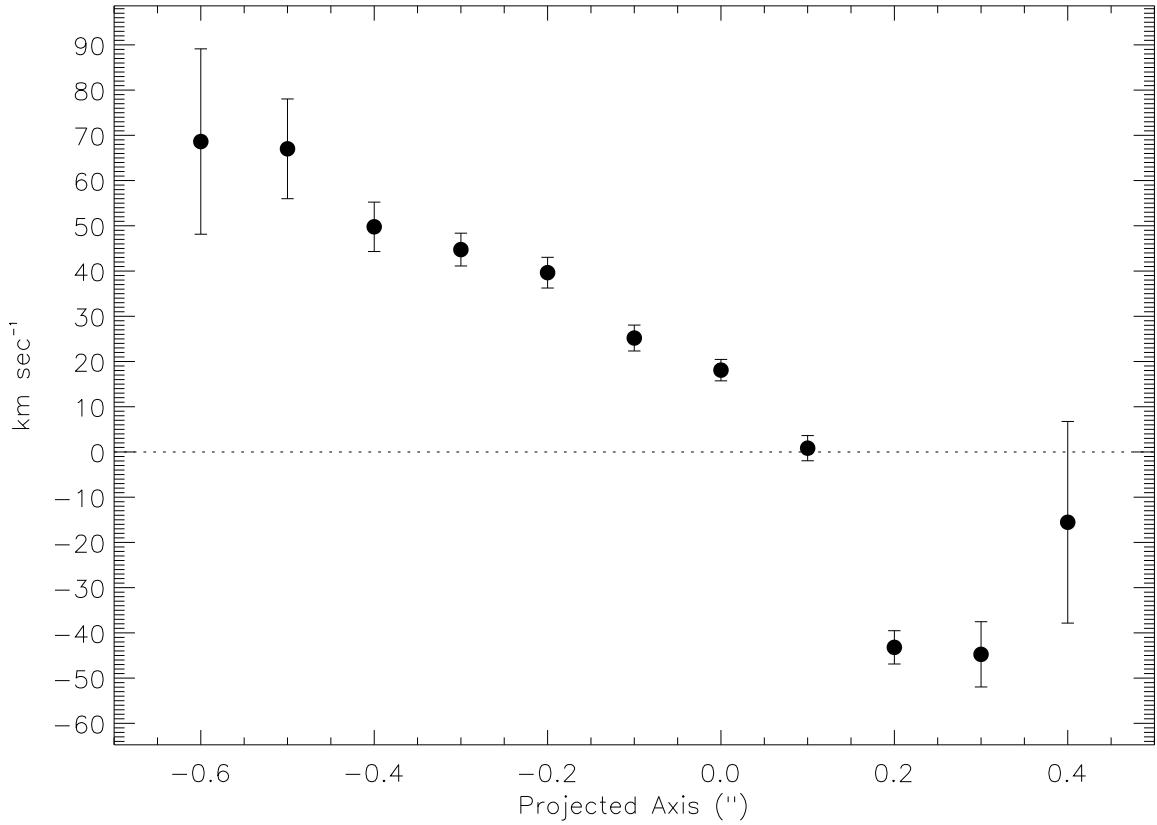}
\caption{Velocity (km s$^{-1}$) versus the projected axis (\arcsec) for the 160$^\circ$ position angle determined from the spider diagram modeling in \S\ref{spider}. This plot represents a slit along the major axis of the galaxy. Projected velocty at each distance was determined from a weighted average of the velocities, using the flux signal-to-noise as weight. The center position (0.0) is defined from the dynamical center of the inclined disk model in  \S\ref{disc}. Errors get larger with greater distance from the center since there were less spatial elements to weight in the average.\label{projang}}
\end{center}
\end{figure*}

To compare with previous slit spectroscopy observations, in Figure \ref{projang} we have plotted the velocity profile for a projected slit across the major axis of the galaxy using our derived position angle of $\phi$=160$^\circ$. Weighted averages of the velocities were determined for each spatial position along the virtual slit using the flux S/N as weight for each velocity position. This illustrates the gradual velocity gradient across the SE portion of the galaxy, and the sharp transition and negative gradient across the dynamical center to the NW region. As seen in previous studies \citep{erb06a,erb06b,schreib06} the \ha~peak is not located at the dynamical center and can be explained by having the \ha~emission arise in discrete SF regions spread through the disk, while the galactic center has a lower state of SF.

Subtracting model velocities from our observed velocities results in residuals from 2 - 35 km s$^{-1}$, with a mean residual of 13 km s$^{-1}$, which is far smaller than the observed residuals of $\sim$170 km s$^{-1}$ from the \citet{genzel06} rotating disk model for their z=2.38 galaxy. Nearby spiral galaxies show typical deviations of $\sim$ 10 km s$^{-1}$ in their velocity profile caused by spiral arms \citep{sof01} and intrinsic dispersions within H$_{II}$ regions \citep{weiner05}. The AO PSF will further smooth the velocity profiles and increase the measured dispersions, since each spatial element of our data cube averages over the AO PSF per wavelength channel and the spectral features are broadened. However, this effect of beam smearing can not be fully determined without knowing the intrinsic distribution of H$\alpha$ emission. We investigate beam smearing effects by simulating a galaxy's \ha~flux distribution of both an exponential disk or two compact sources in an OSIRIS data cube, convolved with varying PSF performances of the AO system. For a Strehl ratio of 20\% close to our projected AO performance, the typical beam smearing on velocity dispersions range from 18 to 33 km sec$^{-1}$ for an exponential disk derived from our modeled inclined disk parameters. For two compact sources with differing intensities we find velocity dispersions from 5 to 55 km sec$^{-1}$, where the peak dispersion occurs at the brightest knot. Both of these models, imply that a significant portion of the observed velocity dispersions are intrinsic to BM133, particularly within the NW region.

The mean internal velocity dispersion for a given spatial location in BM133 is 68 km s$^{-1}$ with the highest $\sigma_{v}$ near the flux center, just NW of the dynamical center. Smaller $\sigma_{v}$ values occur SW from the dynamical center. In \citet{schreib06} much higher dispersions (often greater than 100 km s$^{-1}$) are observed. Although their seeing limited observations likely suffer from significantly more beam smearing, this is unlikely sufficient to explain the differences between the two samples. They use the large ratio of velocity dispersion to circular velocity ($\sigma_{v}$/v$_{c}$) to argue that the disks are temporary and will likely contract to form bulges as they dissipate random motion. Similarly, the single z=2.38 galaxy observed by \citet{genzel06} with AO also shows very high velocity dispersions which are too large to be explained through beam smearing. The best fit disk in \citet{genzel06} also leaves residuals of as much as 170 km s$^{-1}$, implying even stronger velocity components that are not from simple disk rotation. With our observed dispersions of 68 km sec$^{-1}$ and mean residual velocities of 15 km s$^{-1}$ of our simple disk model, we believe that BM 133 is a candidate for a true disk that may persist to the present day. In addition with the ability of LGS-AO we are able to resolve a compact knot region ($\sim$200 pc) in the NW of the galaxy, this may be indicative of high star formation region(s) either within a portion of the disk or the possiblity of an early-stage merger as we discuss in Section \S\ref{add}. We note that its redshift of 1.4781 places it at a look back time of 9.3 Gyr, comparable to the estimated age of the Milky Way Disk of 7.3$\pm$1.5 Gyr from studies of white dwarfs \citep{hans02} and of 7.5$\sbond$10 Gyr from isochrone fitting of the thick-disk cluster NGC 6791 \citep{sand03}.

\subsection{Mass Distribution}\label{mass}

Using our disk model and observed kinematics we are able to investigate the mass distribution of BM133. We calculate an enclosed dynamical mass (M$_{encl}$) of a highly flatten spheriod assuming Keplerian motion,

\begin{equation}
M_{encl} = \frac{2 v_{c}^{2}r}{\pi G} 
\end{equation}

\noindent{For \textit{r}, we choose the farthest distance from the dynamical center with a measured velocity offset at r = 0\farcs64 (5.5 kpc). We use the circular velocity (v$_{c}$ = v$_{p}$ = 72 km s$^{-1}$) determined from the inferred spider diagram, and the gravitional constant (G), to find a dynamical mass of M$_{encl}$ = 4.3$\times$10$^{9}$ \msun.}

The enclosed mass is a lower limit on the total mass of BM133 since it underestimates the halo mass of the galaxy which includes stellar, gas, and dark matter mass components beyond the radius (\textit{r}) assumed. Following \citet{schreib06}, we are able to estimate the halo mass (M$_{halo}$) of BM133 using the spherical virialized collapse model \citep{peebles80, white91}. These models estimate the halo mass with the following equation,
 
\begin{equation}
M_{halo} = \frac{1}{2} \left( \frac{v_{c}}{1.67} \right) ^{3} H_{o}^{-1} G^{-1} (1+z)^{-1.5} 
\end{equation}

\noindent{We find a total halo mass of M$_{halo}$ = 3.4$\times$10$^{10}$ \msun for BM133. This halo mass is roughly an order-of-magnitude lower than the \citep{schreib06} z $\gtrsim$ 2 sample. In addition, we confirm our halo mass by deriving a similar estimate from \citet{erb03} by recognizing that $\Omega_{M}$/$\Omega_{B}$ $\sim$ 7, where $\Omega_{B}$ is the total baryonic matter from M$_{stellar}$ and M$_{gas}$. Therefore, we place a limit on the total halo mass of BM133 to be 7 times larger than its stellar mass, which likely dominates within the \ha~region. With our calculated enclosed mass we find  M$_{halo}$ $\gtrsim$ 4.8$\times$10$^{10}$ \msun, which is in reasonable agreement with our spherical virialized collapsed model estimate. The halo mass derived is a lower limit given that modern galaxies have much smaller ratios of baryonic to dark matter.} 

The virial theorem provides an additional way of estimating the enclosed mass of a galaxy, assuming velocity dispersion is due to the collective motion of the gas and stars in a gravitionally bound system according to the following equation,

\begin{equation}
M_{vir} = \frac{C\sigma_{v}^{2}r_{1/2}}{G}
\end{equation}

\noindent{The galaxy's mass distribution and velocity isotropy influences the constant factor (\textit{C}), which usually ranges from 1 to 10. Most of these contributions are unknown, previous work adopted a uniform spherical distribution of \textit{C}=5 \citep{pett01,erb03,erb04}. For comparison, we follow the \citet{erb06a} assumption of disk geometry and use C = 3.4. With an effective half light radius ($r_{1/2}$) for BM133's \ha~flux of $\sim$0\farcs3 (2.56 kpc), and using the global velocity dispersion we find a virial mass of M$_{vir}$ = 1.1$\times$10$^{10}$ \msun. This is comparable to the \citet{erb06a} estimate of 1.8$\times$10$^{10}$ \msun, assuming their observed $\sigma_{v}$ and their assumed galaxy radius ($\sim$8 kpc). Our virial mass estimate is a factor of 2.6 greater than our estimated enclosed mass. Several factors likely explain this relatively large disagreement. In particular, by using \textit{C}=3.4 we are assuming an inclination of 45$^{\circ}$ instead of our model's value of \textit{i}=54$^{\circ}$. More importantly, the dispersions are likely significantly higher than the model assumes due to the irregular distribution of H$\alpha$ flux being a biased tracer of the random motion. In addition, we can not rule out a merger scenario where two knot-like structures are infalling, and therefore induce high SF in each merger component. A possible merger scenario is further discussed in \S\ref{add}. } 

Of all the mass estimates we favor the enclosed mass from the inclined-disk model since it includes inclination effects and uses intrinsic circular velocities. However, we recognize that this enclosed mass is substantially lower than other mass estimates of star forming galaxies at this redshift range. Our mass estimates are probing the inner kiloparsecs of the galaxy, and BM133 is likely to be imbedded in a much larger, extended structure beyond our observed radius. We note that \citet{erb06a} find a gas fraction mass of $\sim$85\% giving BM133 one of the highest star formation rates per unit stellar mass in the NIRSPEC sample. This large gas fraction will likely lead to further contraction which will steepen its velocity profile and produce larger velocities overall. Therefore the slope of the observed velocity profile is a lower limit to the eventual modern value. The kinematics observed still fit within the range of rotation curves found in nearby spiral galaxies with comparable enclosed massses out to these radii (ie, NGC 598 : \citealt{sof99,sof01}). However, we stipulate that BM133 may continue to grow into a more massive disk system given its observed youth and gas richness, where the majority of the growth will occur at larger radii. 

The enclosed, virial, and halo masses inferred here are model dependent and are extrapolations from the observed \ha~distribution and kinematics which are concentrated near the center of the galaxy. We investigate these quantities in part to compare with prior mass estimates of BM133 \citep{erb06a} and mass estimates of similar objects in the \citet{schreib06} survey. A significant difficulty in such a comparison is that the IFS LGS-AO observations focus on the inner several kpc, while the seeing-limited campaigns extend to larger radii, albeit with poorer resolution. Therefore the mass estimates will likely represent a lower limit since we do not probe the entire distribution of even the baryon matter of the galaxy.

\subsection{Metallicity}\label{met}

We did not detect optical rest-frame [N{\sc ii}] emission, however we were able to place an upper limit to [N{\sc ii}]/\ha~$\lesssim$ 0.12. With such a low [N{\sc ii}]/\ha~ratio we can rule out that BM133 is dominated by an active galacic nucleus \citep{bald81}. Using the relationship between oxygen abundance (O/H) and [N{\sc ii}]/\ha~ previously used in LBG studies \citep{pett01,shap01,pett04} we can estimate the H$_{II}$ region abundances. With this relationship we find an oxygen abundance upper limit of 12 + log(O/H) $\lesssim$ 8.38. The assumed oxygen abundance for solar metallicities is 12 + log(O/H) = 8.65 \citep{pett04}, therefore we infer a metallicity less than 1/2 solar for BM133. Our [N{\sc ii}]/\ha~limit is consistent with other abundance studies for z$\gtrsim$2 LBG samples \citep{shap01,erb03,erb06c,schreib06}. Our inferred (O/H) abundance limit is in reasonable agreement with \citet{erb06c} relantionship between (O/H) abundance versus stellar mass, assuming their stellar mass of 5x10$^{9}$ \msun from SED fitting.

\subsection{Possible Merger Scenario}\label{add}

Although a smooth disk-like velocity structure can explain the observed kinematics, we note that with the LGS-AO resolution, much of the H$\alpha$ flux is found in a single compact region ($\sim$200 pc) to the NW of the kinematic center. It is difficult to determine if this is a region of extremely high star formation within a disk, or a separate merging galaxy. In either case, the intensity of star formation in such a small region is extreme with an observed SFR $\gtrsim$ 16 \msun yr$^{-1}$, and the region shows the highest internal dispersion within all of BM133. If this is a star forming region within an otherwise stable disk, then it represents an extreme mode of star formation. But in this gas rich system it might be possible to temporarily support such run-away star formation. It seems reasonable that such a region would have significantly higher internal turbulence compared to modern star forming regions helping to explain the high intrinsic dispersion.

In order to investigate this scenario we measure velocity dispersions ($\sigma$$_{v}$) for two 0\farcs3 $\times$ 0\farcs3 regions NW and SE of the galaxy. Assuming that each of these components reflect the internal dispersion of gas and stars and have a half-light radius of r$_{1/2}$ = 0\farcs1 (0.9 kpc), we find a virial mass for each component of M$_{vir, NW}$ = 6.5$\times$10$^{9}$ \msun and M$_{vir, SE}$ = 2.3$\times$10$^{9}$ \msun. If both of these components are infalling towards each other, we can estimate the expected observed velocity (v$_{obs}$sin \textit{i}) by considering Keplerian motion with the given separation (3.81 kpc) and virial masses. We find that for such a merger system the observed velocity would have an upper limit of 80 km s$^{-1}$. This velocity is quite similar to our observed plateau velocity of v$_{p}$ = 72 km s$^{-1}$, therefore we can not rule out the possibility of a merger scenario. However, if this was a merger it would also be fortuitous to have tidal stripping between the two components that would produce such a smooth velocity gradient with a dispersion highest near one of the flux centers as we observe in BM133. So, although we cannot fully rule out an early-stage merger system, we believe that it is a less probable scenario given our measured two-dimensional velocity offsets and dispersions, and our well-fit inclined-disk model.

\section{Summary}\label{sum}

We have presented first extragalactic results from the new integral field spectrograph OSIRIS with the Keck LGS AO system by measuring \ha~emission from a z=1.4781 star forming galaxy. We spatially resolve \ha~over a 6.8 $\times$ 4.3 kpc region and find a global flux of 4.2$\pm$0.6$\times$10$^{-16}$ ergs s$^{-1}$. We find a high dereddened SFR of 66$\pm$9 \msun yr$^{-1}$. Based on [N{\sc ii}]/\ha~ratio limit we infer a 1/2 solar or below metallicity, which is consistent with other high-z LBG samples. We fit an inclined-disk model to the 2D kinematics of BM133, and find low residuals with a disk having a position angle ($\phi$) of 160$^{\circ}$, inclination angle (\textit{i}) of 54$^{\circ}$, and plateau velocity of 72 km s$^{-1}$ achieved within a radius of $\sim$ 0\farcs11 (0.9 kpc). We estimate an enclosed mass of 4.3$\times$10$^{9}$ \msun and virial mass of 1.1$\times$10$^{10}$ \msun, which reconfirms results by \citet{erb06a,erb06b}. We cannot conclusively eliminate a merger scenario, but we favor a true disk in an early stage of star formation. 

An IFS combined with LGS-AO offers a powerful new tool for studying morphology and kinematics of z $\gtrsim$ 1 galaxies. The capability to achieve diffraction-limited NIR observations, comparable to optical HST observations, offers a superb and necessary capability for studying kinematics and evolution of high-z galaxies. Although results presented in this paper are for only one target, the analysis demonstrated here is being applied to a larger on-going study of star forming galaxies at these redshifts. The capability of near-IR IFS instruments such as OSIRIS and SINFONI to provide 2D mapping of optical rest-frame spectral features of high-z galaxies has already demonstrated significant advantages for investigating kinematics and mass distribution of these objects. BM133 has offered a stepping stone for future high-z extragalactic kinematic studies, and offers one of the best candidates for a precursor of present-day spiral galaxies.

\acknowledgements
The authors would like to sincerely thank the dedicated members of the Keck Observatory staff (CARA) who contributed greatly to the success of the commissioning of OSIRIS. We would like to acknowledge our appreciation to Sean Adkins, Paola Amico, Randy Campbell, Al Conrad, Allan Honey, David Le Mignant, Jim Lyke, Marcos van Dam, and Peter Wizinowich. Data presented herein were obtained at W.M. Keck Observatory, which is operated as a scientific partnership between the California Institute of Technology, the University of California and the National Aeronautics and Space Administration. The Observatory was made possible by generous financial support of the W.M. Keck Foundation. The authors wish to recognize and acknowledge the significant cultural role and reverence that the summit of Mauna Kea has always had within the indigenous Hawaiian community.  We are most fortunate to have the opportunity to conduct observations from this ``heiau" mountain.

\clearpage
\begin{deluxetable}{lccccccccc}
\label{obs}
\tabletypesize{\scriptsize}
\tablecaption{Q2343-BM133 Observational Properties}
\tablewidth{0pt}
\tablehead{
\colhead{RA} & 
\colhead{DEC} & 
\colhead{z$_{H\alpha}$} & 
\colhead{K$_{s}$\tablenotemark{\dagger}} &
\colhead{Exposure} & 
\colhead{Date} & 
\colhead{F$_{H\alpha}$} & 
\colhead{[NII]/\ha} & 
\colhead{$\Delta$v} & 
\colhead{$\sigma_{v}$} \\ 
\colhead{(J2000)} & 
\colhead{(J2000)} & 
\colhead{} & 
\colhead{(mags)} & 
\colhead{(seconds)} &
\colhead{Observed} &
\colhead{(ergs s$^{-1}$ cm$^{-2}$)} & 
\colhead{} & 
\colhead{(km s$^{-1}$)} & 
\colhead{(km s$^{-1}$)} \\
}
\startdata
23:46:16.18 & 12:48:09.31 &  1.4781 & 20.5 & 6x900 & November 2005  & 4.2$\pm$0.6x10$^{-16}$ & $\textless$ 0.12 & 134  & 73$\pm$9 \\
\enddata
\tablenotetext{\dagger}{Observed by \citet{erb06a,erb06b}}
\end{deluxetable}

\begin{deluxetable}{lccccccccc}
\label{props}
\tabletypesize{\scriptsize}
\tablecaption{Q2343-BM133 Derived Properties}
\tablewidth{0pt}
\tablehead{
\colhead{SFR} & 
\colhead{SFR$_{dered}$} &
\colhead{r$_{1/2}$} &
\colhead{v$_{p}$\tablenotemark{\dagger}} & 
\colhead{r$_{p}$} &
\colhead{Position Angle} & 
\colhead{Inclination} & 
\colhead{M$_{encl}$} &
\colhead{M$_{vir}$} &
\colhead{M$_{halo}$} \\ 
\colhead{(\msun yr$^{-1}$)} &
\colhead{(\msun yr$^{-1}$)} &
\colhead{(kpc)} &
\colhead{(km s$^{-1}$)} &  
\colhead{(kpc)} & 
\colhead{($\phi$)} & 
\colhead{(\textit{i})} & 
\colhead{(\msun)} &
\colhead{(\msun)} &
\colhead{(\msun)} \\
}
\startdata 
47$\pm$6 & 66$\pm$9 & 2.6 & 72 & 0.9 & 160 & 54 & 4.3x10$^{9}$ & 1.1x10$^{10}$ & 3.4x10$^{10}$ \\
\enddata
\tablenotetext{\dagger}{v$_{p}$ is used for v$_{c}$ in the enclosed and halo mass estimates}
\end{deluxetable}

\clearpage

\end{document}